\documentclass[twocolumn, deluxetables]{aastex631}
\usepackage{amsmath}
\usepackage[caption=false]{subfig}
\usepackage{multirow}
\usepackage{graphicx}
\usepackage{xcolor}
\usepackage[T1]{fontenc} 
\usepackage{enumitem}
\usepackage{comment}
\usepackage{CJK}

\newcommand{\xmm}{\textit{XMM-Newton}}

\shorttitle{XMM Mrk 877}
\shortauthors{Xiang et al.}

\graphicspath{{./}{figures/}}

\begin{document}
\begin{CJK*}{UTF8}{gbsn}

\title{Multi-layered Fast Wind observed in XMM-Newton snapshot of Seyfert 1 Markarian 877}

\author[0000-0002-7129-4654]{Xin Xiang}
\affiliation{Department of Astronomy,
The University of Michigan, 1085 South University Avenue, Ann Arbor, Michigan,48103, USA}
\email{xinxiang@umich.edu}

\author[0000-0003-2869-7682]{Jon M. Miller}
\affiliation{Department of Astronomy,
The University of Michigan, 1085 South University Avenue, Ann Arbor, Michigan,48103, USA}

\author[0000-0001-9735-4873]{Ehud Behar}
\affiliation{Department of Physics, Technion, Technion City, Haifa 3200003, Israel}

\author[0000-0002-0167-2453]{W. N. Brandt}
\affiliation{Department of Astronomy \& Astrophysics and the Institute for Gravitation and the Cosmos, The Pennsylvania State University, 525 Davey Lab, University Park, PA}

\author[0009-0006-4968-7108]{Luigi Gallo}
\affiliation{Department of Astronomy and Physics, Saint Mary's University, Nova Scotia B3H 3C3, Canada}

\author[0000-0002-3687-6552]{Doyee Byun}
\affiliation{Department of Astronomy,
The University of Michigan, 1085 South University Avenue, Ann Arbor, Michigan,48103, USA}

\author[0000-0001-5802-6041]{Elena Gallo}
\affiliation{Department of Astronomy,
The University of Michigan, 1085 South University Avenue, Ann Arbor, Michigan,48103, USA}


\begin{abstract}
Ultra Fast Outflows (UFOs) are powerful, highly ionized winds launched from the innermost regions of Active Galactic Nuclei (AGNs), reaching velocities of 0.03 -- 0.3 c and playing a key role in AGN feedback. We present a photoionization analysis of an 18 ks \xmm\ snapshot of the Seyfert 1 AGN Mrk 877, revealing three distinct UFO components with line-of-sight velocities of $0.10^{+0.005}_{-0.005}~c$ , $ 0.04^{+0.005}_{-0.004}~c$ , and $0.05^{+0.005}_{-0.004}~c$. These components span a broad range of ionization parameters and column densities, producing absorption features across both soft and hard X-ray bands. Even under the most conservative assumption for the volume filling factor, the fastest component exceeds $5\%$ of the Eddington luminosity, making it capable of driving strong galaxy-scale feedback. The soft X-ray UFO component, despite its lower ionization, shares a similar velocity as a higher-ionization component, hinting at a two-phase medium likely shaped by clumpiness or interactions with ambient material.  The density profile inferred from the Absorption Measurement Distributions (AMD) and the positive trend between outflow momentum rate and radiation momentum flux suggest that wind is powered by a combination of radiative and magnetic driving.

\end{abstract}


\section{Introduction} \label{sec:intro}
Powerful, ionized winds from Active Galactic Nuclei (AGNs), the rapid phases of supermassive black hole (SMBH) growth, are widely recognized as one of the most effective feedback mechanisms in galaxy evolution \citep{King_Pounds_2015}. If the kinetic power of such outflows exceeds 0.5--5\% of the Eddington luminosity, they are capable of significantly impacting the host galaxy interstellar medium and star formation \citep{Hopkins_2010, Di_2005}. Probing the origin and energetics of these winds, including their launching radii, driving mechanisms, and total feedback, is essential for understanding the co-evolution of SMBHs and their host galaxies.

Ultra-fast outflows (UFOs) are powerful, highly ionized winds launched from the innermost regions of AGNs, reaching velocities of $0.03-0.3c$ \citep{Chartas_2002, Tombesi_2011, Gofford_2015, Igo_2020, Gallo_2023}. They are observed in both near-Eddington sources (e.g. \citealt{Nardini_2015, Zoghbi_2015, Reeves_2018}) and sub-Eddington sources (e.g. \citealt{Zak_2024, Parker_2017, Xiang_2025, Mehdipour_2025}). 

Highly ionized UFO components were observed as relativistic blue-shifted Fe K-shell absorption lines from Fe XXV (He-like, 6.70 keV) and Fe XXVI (H-like, 6.97 keV) at E $>$ 7 keV, with ionization parameter in the range of $\log{\xi} \sim 3-6~\mathrm{erg~s^{-1}~cm}$, and column density between $N_H \sim 10^{22}-10^{24} \mathrm{cm^{-2}}$ \citep{Tombesi_2011, Igo_2020, Gofford_2013}. The ionization parameter is defined by $\xi = L_{\mathrm{ion}}/n r^2$ \citep{Tarter_1969} where $L_\mathrm{ion}$ is the ionizing luminosity of the source in the energy band 13.6 eV - 13.6 keV, $n$ is the electron number density of the gas, and $r$ is the distance from the SMBH. UFO components in soft X-rays were revealed via L-shell transitions from highly ionized Fe, and lines from highly ionized O, Ne, Si, and Mg, with lower ionization parameters around $\log{\xi} \sim 1-3~\mathrm{erg~s^{-1}~cm}$ \citep{Reeves_2018, Reeves_2003}. 

Since kinetic power in an outflow is a steep function of velocity ($\dot{K}_E \propto v^3$), UFOs can carry significant energy, making them compelling candidates for AGN feedback -- one of the leading mechanisms proposed to regulate the co-evolution of galaxies and their central SMBHs \citep{King_Pounds_2015, Tombesi_2010, Gofford_2015}.

The physical mechanisms driving these outflows likely vary with the accretion state of the AGN. In sources accreting near or above the Eddington limit, radiation pressure via electron scattering may suffice to launch UFOs \citep{Nardini_2015, Reeves_2018}, as is suggested for PDS 456 \citep{XRISMPDS456_2025}. However, for sub-Eddington sources, UFOs must be driven through other processes: radiation pressure on lines \citep{Castor_1975, Dannen_2019} or magnetic pressure \citep{Blandford_Payne_1982, Contopoulos_Lovelace_1994} or a combination of both \citep{Proga_2003, Everett_2005}, likely the case of NGC 4151 \citep{Xiang_2025} and NGC 3783 \citep{Mehdipour_2025}. Each of these mechanisms predicts distinct density profiles and ionization structures \citep{Behar_2009}, which can be tested through correlations between column densities, ionization parameters, and outflow velocities.

Mrk 877 (PG 1617+175) is a nearby ($z = 0.11$) radio-quiet type 1 quasar (high-luminous Seyfert 1) with no significant optical obscuration \citep{Jang_1995} with a bolometric luminosity of $L_{bol}\sim 5\times10^{45} \mathrm{erg~s^{-1}}$ \citep{Veron_2006}. The black hole mass has been estimated via reverberation mapping to be $M_{BH} = 5.2^{+1.1}_{-1.3} \times 10^8 M_{\odot}$ \citep{Bentz_2013, Batiste_2017}. Early optical spectroscopy showed pronounced red asymmetry of its broad H$\beta$ profile, possibly arising from the BLR \citep{De_1985}. HST/COS data \citep{Veilleux_2022} reveals broad, blueshifted absorption troughs with velocities $\sim 3000~\mathrm{km~s^{-1}}$ in N \textsc{v}, O \textsc{vi}, and especially P \textsc{v}, which requires large column densities and high ionization parameters. These features indicate a substantial mass of fast, highly ionized gas consistent with a disk wind launched close to the accretion disk, directly linking small-scale to large-scale outflows. Despite its brightness and larger-scale outflow features, Mrk 877 has received little attention in the X-ray band, leaving its inner disk wind properties largely unexplored.

This paper presents a detailed X-ray spectral analysis of Mrk 877 using an archival observation from \xmm\ in 2018. The spectra reveal multiple layers of UFO absorbers with velocities exceeding $0.03c$, spanning a wide range of ionization and column densities. Section~\ref{sec:data} describes the observations and data reduction. Section~\ref{sec:results} details the spectral fitting procedure and derived absorber properties. In Section~\ref{sec:discussion}, we summarize our findings and discuss their implications, the limitations of our analysis, and future perspectives.

\section{Observation and Data Reduction} \label{sec:data}
Mrk 877 was observed with XMM-Newton through ObsID 0801892401, starting on 12 March 2018 at 04:44:31 UTC.  The nominal exposure time was 24~ks.  Using SAS version 21.0.0 and the associated calibration files, we reduced the data from the EPIC cameras and the RGS.  Mrk 877 is only moderately bright, so we elected to concentrate on data from the EPIC-pn camera owning to its broad energy range and high sensitivity.  Using blank regions on the same CCD as Mrk 877, we generated background light curves and time filters to exclude flaring events.  The event list was additionally filtered to only include event grades 0--4, and those with ``FLAG=0'' as per standard procedures.  After these filtering steps, the net EPIC-pn exposure time was 18.2~ks.  Source and background regions with radii of 24 arc-seconds were selected to create spectra.  Channels between 0-20479 were binned by a factor of 5.0, again as per standard practice.  The SAS tools ``arfgen'' and ``rmfgen'' were used to create responses.

\section{Analysis and Results} \label{sec:results}
In this section, we present a spectral analysis of the observation data in the 0.3-10 keV range. The fits presented were performed with SPEX version 3.08.01 \citep{Kaastra_1996}, minimizing a Cash statistic (C-stat; \citealt{Cash_1979}). The uncertainties for parameters reported in this work are $1\sigma$ errors. In order to increase the signal-to-noise, binning factors of 3, 7, 12, 15, and 30 are applied to the 0.3--1, 1--3, 3--5, 5--8, and 8--10 keV energy ranges, respectively.

\subsection{Modeling procedures}

\begin{figure*}
\includegraphics[width=\textwidth]{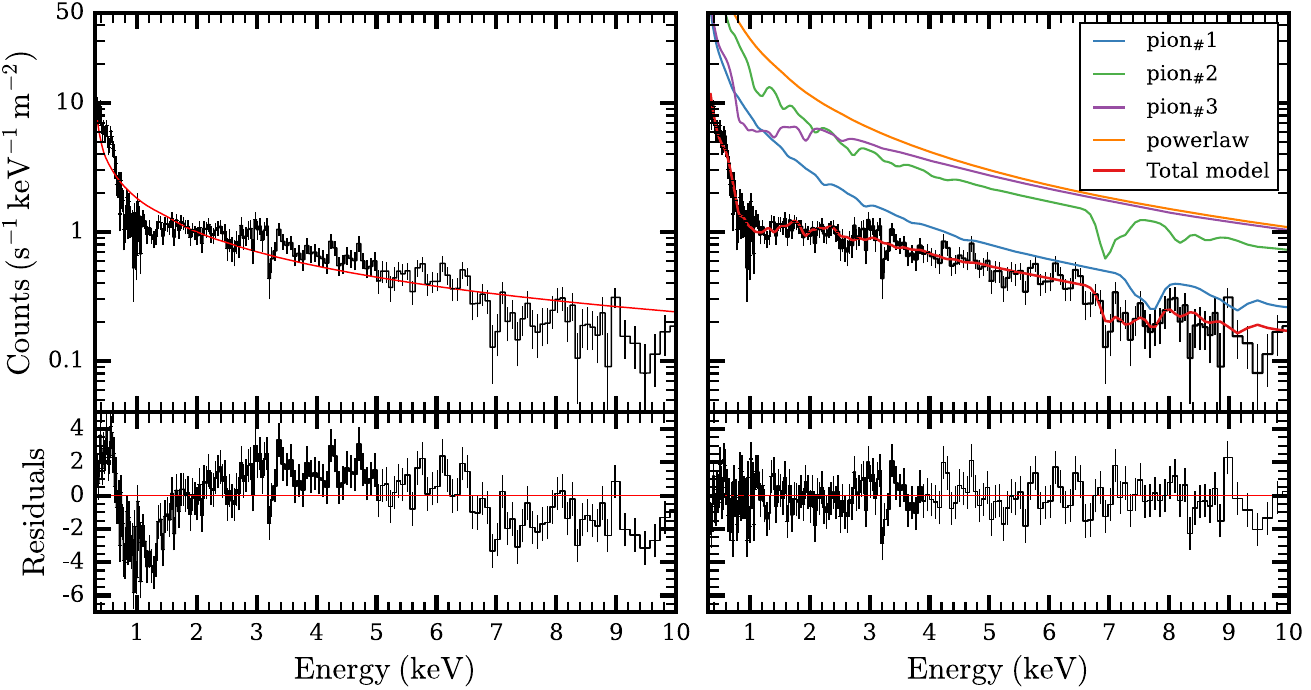}
\caption{LEFT: XMM-Newton/PN 18 ks snapshot of Mrk 877 in the source frame, fit with a simple power-law. The strong flux deficits between 0.7--1.4 keV and 6.4--8.4 keV indicate multiple ionized partially covering absorption components. RIGHT: Best-fit model using three distinct UFO components with the \texttt{$pion$} photoionization model in SPEX. The green, magenta, and blue curves represent absorption from low- to high-ionization zones. These components are detected at $5\sigma$, $4\sigma$, and $> 10\sigma$. The red curve shows the total best-fit model.}
\label{fig: pl_vs_pion3}
\end{figure*}

The left panel of Figure \ref{fig: pl_vs_pion3} shows a simple power-law fit that yields a C-stat/d.o.f = 837/187 and reveals strong residuals between 0.7--1.4 keV and 6.4--8.4 keV, indicative of absorption features. The power-law captures the broad spectral shape, consistent with the typical continuum expected from a Seyfert 1. The absorption features are likely contributed from outflowing ionized gas, similar to the cases of PG 1211+143 and Mrk 817 \citep{Reeves_2018, Zak_2024}. To model these features self-consistently, we employed the ``\texttt{$pion$}'' photoionization model in SPEX \citep{Miller_2015, Mehdipour_2016}. The ``\texttt{$pion$}'' model calculates the ionization balance from the incident ionizing radiation from the inner component, which provides a physically realistic treatment of layered wind components. To include the important FUV contribution to the ionizing luminosity for estimating the ionization parameters for the absorption features, a blackbody component ``\texttt{$bb$}'' with a disk temperature of 0.01 keV and emitting area of $2\times10^{25}~\mathrm{m^2}$ at peak emissivity is included, which is appropriate estimates for the standard accretion disk \citep{Shakura_1973} with central SMBH mass of $M_{BH} \sim 5\times10^8 M_{\odot}$. To ensure a realistic spectral shape and ionizing luminosity, the power-law component ``\texttt{$pow$}'' bends to zero flux at both low (``\texttt{$etau_{low}$}'') and high (``\texttt{$etau_{hi}$}'') energies with cutoffs of 0.0136 keV and 300 keV. The redshift model ``\texttt{$reds$}'' is applied with a fixed source redshift of $z = 0.11$, shifting all the emission and absorption components to the frame of Mrk 877.

The treatment of ``\texttt{$pion$}'' model is similar to the photoionization modeling presented in \cite{Xiang_2025}. Five parameters from the ``\texttt{$pion$}'' absorption components are left free: (1) $N_H~(\mathrm{cm^{-2}})$, the column density; (2) $\xi~(\mathrm{erg~cm~s^{-1}})$, the ionization parameter; (3) $v_z (\mathrm{km~s^{-1}})$, the bulk velocity of the gas in the source frame along our line-of-sight (negative for blue-shifted components); (4)$\sigma_z~(\mathrm{km~s^{-1}})$, the turbulent velocity of the gas; and (5) $f_{cov}$, the geometric covering factors for the absorbers defined as $f_{cov} = \Omega/4\pi$ ($\Omega \in [0, 4\pi]$), where $f_{\mathrm{cov}} = 1$ represents a spherical shell of gas. Other ``\texttt{$pion$}'' parameters retained their default values. The elemental abundances were kept at unity relative to solar values in all fits \citep{Lodders_2009}. The ionizing luminosity $L_{\mathrm{ion}}$ of each ``\texttt{$pion$}'' layer is automatically calculated by the model.

After a detailed assessment, three layers of ``\texttt{$pion$}'' are statistically required to fully explain the absorption features. The total best-fit model can be summarized as follows:
\begin{multline}
     (bb + (pow * etau_{low} * etau_{hi})) * \\  pion_{\#1} * pion_{\#2} * pion_{\#3} * reds
\end{multline}
The right panel of Figure \ref{fig: pl_vs_pion3} shows the best fit model, with a breakdown of the individual ``\texttt{$pion$}'' components. This model represents a significant improvement over the simple power-law fit and captures all the prominent residuals present in the data. With the three ```\texttt{$pion$}''' components included, the C-stat is reduced to $C = 239$ for 172 degrees of freedom ($\nu$). The expected C-stat for this number of degrees of freedom is approximately $C_e = 190$, with a $1\sigma$ range of $\sqrt{2 \nu} = 19$ \citep{Kaastra_2017}. The observed C-stat is 1.26 times higher than expected, suggesting an acceptable fit. Table \ref{table:parameters} presents the best-fit values and errors for the free parameters in the model as well as some derived properties, including launching radii, mass outflow rates, and kinetic powers (refer to section \ref{sec:prop} and \ref{sec:derivedprop}). The best-fit model gives a total ionizing luminosity of $L_{ion} = 2.9\times10^{45}~\mathrm{erg~s^{-1}}$ and a total obscured flux of $F_{0.3-10} = 2.9 \times 10^{-12} ~\mathrm{erg~cm^{-2}~s^{-1}}$ in the 0.3-10 keV band.

The significance of each wind component is evaluated by removing it from the full model, refitting, and examining the difference in the fit statistic. The corresponding detection significance (D.S) in units of Gaussian-equivalent sigma, derived from the p-value associated with the changes in C-stat and the difference in degrees of freedom. In addition, we compute the Small Sample Akaike Information Criterion (AIC) \citep{Akaike_1974, Emmanoulopoulos_2016}, defined as $\rm AIC = 2\,p - 2\,C_L + C + 2\,p(p+1)/(n-p-1)$, where $C$ is the C-stat of the fit, $C_L$ is the C-stat of the best fit model, $n = 189$ is the number of data bins, and $p = 17$ is the number of free parameters in the model. A difference of 2 or more AIC units indicates substantial support for the model with the lower AIC \citep{Burnham_2002}.  In Table~\ref{table:parameters}, we report the change in AIC ($\Delta \rm AIC$), calculated by subtracting the AIC of the best-fit model from that of the model with a specific component removed. A value of $\Delta \rm AIC < -10$ indicates that the component is strongly required to account for the spectral features, while $\Delta \rm AIC < -2$ provides substantial evidence.

In addition to the absorption features, we have explored the apparent absence of the neutral Fe K$\alpha$ emission line at 6.4 keV. We included a Gaussian line component (\texttt{gaus}) in the model, centered at 6.4 keV, regardless of whether a visible feature is present in the spectrum. The inclusion of this component results in an identical C-stat value of 239.5, suggesting no significant statistical improvement to the fit. We allowed two parameters to vary freely: the line normalization and its full width at half maximum (FWHM). The best-fit values are poorly constrained, with normalization of $1^{+55}_{-1}\times10^{50}~\mathrm{ph~s^{-1}}$ and FWHM of $17^{+1\times10^{23}}_{-17}~\mathrm{eV}$. Despite the large uncertainties, the normalization allows us to estimate an upper limit on the Fe K$\alpha$ line luminosity. Using the $1\sigma$ upper bound on the normalization, we derive an upper limit of $6 \times 10^{42}~\mathrm{erg~s^{-1}}$ for the line luminosity. This is consistent with the expectations from the observed 3–10 keV continuum luminosity of $4 \times 10^{44}~\mathrm{erg~s^{-1}}$ \citep{Waddell_Gallo_2022}. The absence of the neutral emission line could either be from obscuration or from poor sensitive with the current data. Future observations can distinguish both possibilities.

\subsection{Directly Measured Properties of Wind Components: \texttt{$pion_{\#1}$}, \texttt{$pion_{\#2}$}, and \texttt{$pion_{\#3}$}} \label{sec:prop}

\begin{figure*}
\includegraphics[width=\textwidth]{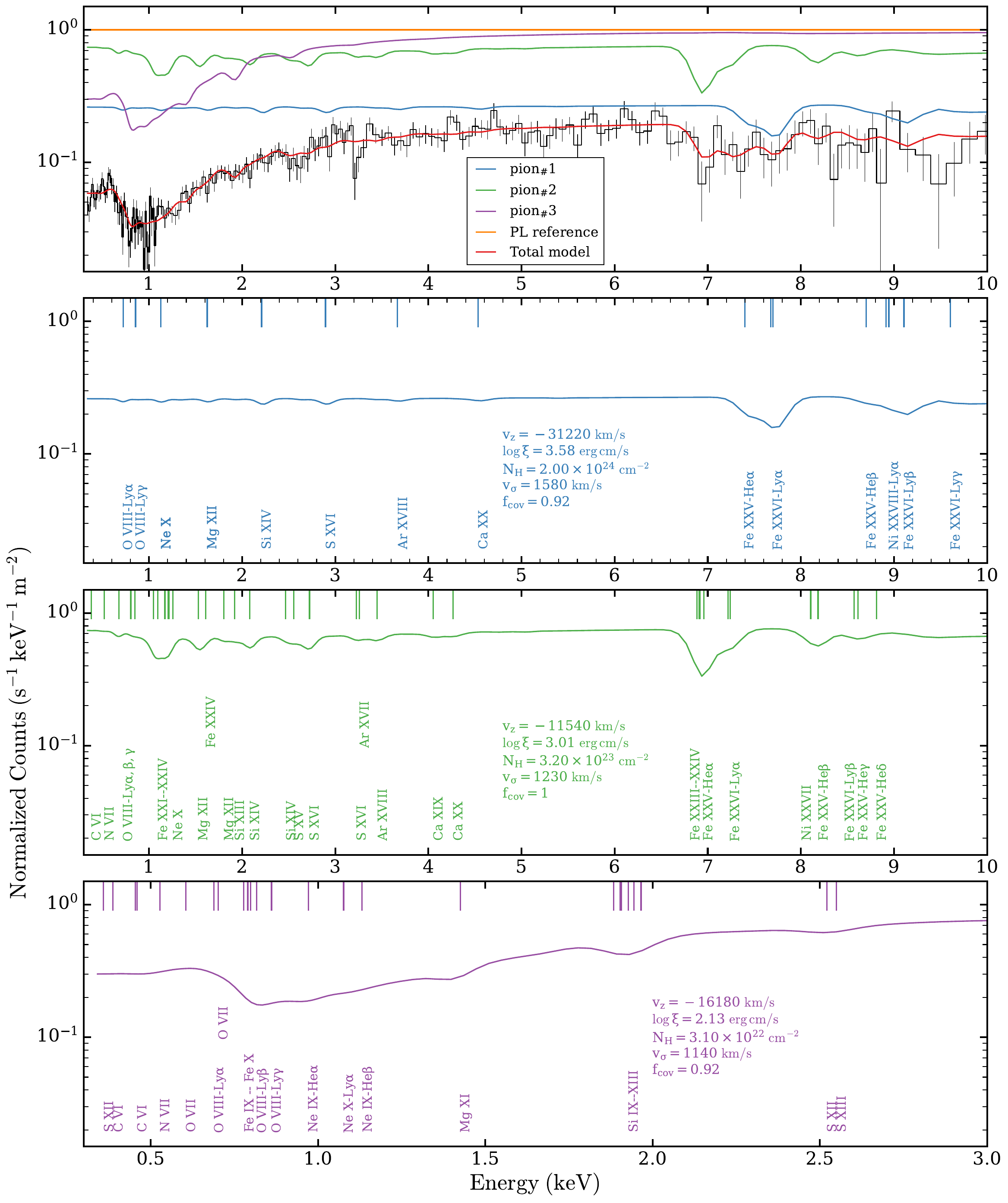}
\caption{Top Panel: The observed X-ray spectrum of Mrk 877 with the best-fit model (red curve), plotted relative to the underlying power-law continuum (orange curve) to highlight the absorption features. The contributions from each of the three ``\texttt{$pion$}'' components are shown separately by blue, magenta, and green curve, respectively. Bottom Three Panel: The transmission models of each ``\texttt{$pion$}'' component plotted individually, with vertical lines marking the most prominent absorption lines from different elements. The bottom panel for ``\texttt{$pion_{\#3}$}'' is plotted in a range of 0.3 -- 3 keV, while all other panels are plotted in a range of 0.3 -- 10 keV.}
\label{fig:3pion}
\end{figure*}

The wind component, \texttt{$pion_{\#1}$}, that is layered closest to the central engine in the SPEX modeling configuration, has the most extreme properties. It has an LOS outflow velocity of $v_z = -31220^{+1270}_{-1450}~\mathrm{km~s^{-1}} \sim 0.1~c$, a large column density of $N_H = 2.0^{+0.3}_{-0.4} \times 10^{24}~\mathrm{cm^{-2}}$, and a high ionization parameter $\log \xi = 3.58^{+0.14}_{-0.12}~\mathrm{erg~cm~s^{-1}}$. This component dominates the absorption in the Fe K band, producing prominent features from highly blueshifted Fe XXV-He$\alpha$ (6.70 keV), Fe XXVI-Ly$\alpha$ (6.97 keV), and Fe XXVI-Ly$\beta$ (8.25 keV). The inclusion of $pion_{\#1}$ is statistically significant: the best-fit model without it has a C-stat increased by 49, while reducing the degrees of freedom by 5. This corresponds to a $6 \sigma$ detection and a $\Delta AIC$ of -37, indicating that this component is strongly required by the data.

The second wind layer, \texttt{$pion_{\#2}$}, represents an intermediate-ionization UFO absorber with $v_z = -11540^{+1410}_{-1070}~\mathrm{km~s^{-1}} \sim 0.04~c$,
$\log \xi = 3.01^{+0.05}_{-0.03}~\mathrm{erg~cm~s^{-1}}$, and $N_H = 3.2^{+3}_{-2} \times 10^{23}~\mathrm{cm^{-2}}$. This component produces a dense forest of absorption lines in both soft and hard X-ray bands. Notable transitions include Fe XXI -- Fe XXIV, Mg XII, Si XIV, S XVI, Ar XVII in soft band, and Fe XXV/XXVI in hard band. Its detection is also statistically significant at $4\sigma$ level, with a corresponding $\Delta AIC = -11$.

The outermost layer, \texttt{$pion_{\#3}$}, has a similar outflow velocity of $v_z = -16180^{+1620}_{-1300}~\mathrm{km~s^{-1}} \sim 0.05~c$, but a lower ionization parameter of $\log \xi = 2.13^{+0.03}_{-0.02}~\mathrm{erg~cm~s^{-1}}$, and a lower column density of $N_H = 3.1^{+0.2}_{-0.2} \times 10^{22}~\mathrm{cm^{-2}}$. Its primary contribution to the absorption features in the soft band below 2 keV comes from O VIII, Si IX-XIII, Fe IX, Fe X, S XII, S XIII, and Mg XI. This component is the most statistically significant, with a detection at $>10\sigma$ confidence and a strongly favorable $\Delta AIC = -213$. 

Table \ref{table:parameters} and Figure \ref{fig:3pion} summarize the best-fit parameters and the detailed absorption features of each wind component. Figure \ref{fig: propoties_vs_nH} displays the wind properties as a function of ionization parameter $\log \xi$. Among the plotted quantities, the column density ($N_H$) shows a clear positive correlation with ionization, while the other properties do not exhibit a significant trend. The Absorption Measurement Distribution (AMD) provides insight into the radial density profiles ($n(r) \propto r^{\alpha}$) of the outflows, assuming all outflow layers are produced via the same mechanism or the same ratio of mechanisms at all radii \citep{Behar_2009, Holczer_2007}. Given $\mathrm{AMD} \equiv dN_{\mathrm{H}} / d \log{\xi} \propto \xi^m$ and assuming a power-law distribution of column density as a function of ionization parameters, the AMD slope, m, is directly related to the density profile exponent as $\alpha=(1+2 m)/(1+m) \pm \Delta m / (1+m)^2$. To estimate the slope and account for measurement uncertainties in both $N_{\mathrm{H}}$ and $\xi$, we use the \texttt{emcee} \citep{Foreman-Mackey_2013} Bayesian linear regression. From the posterior, we find a best-fit AMD slope of $m = 1.18 \pm 0.18$, corresponding to a density profile index of $\alpha = 1.54 \pm 0.04$, as shown in the top panel of Figure \ref{fig: propoties_vs_nH}. This is shallower than the assumed $n \propto r^{-2}$ expected for radiatively driven winds \citep{Kazanas_2012}, and is consistent with the prediction for magnetically driven Blandford-Payne (BP) winds ($\alpha = 1.5$; \citealt{Blandford_Payne_1982}). A similar slope was also found in several epochs of XRISM observations on NGC 4151 \citep{Xiang_2025}

To assess the sensitivity of our outflow results to the assumed blackbody contribution, we tested two additional models in which the ``\texttt{$bb$}'' normalization was varied by a factor of 4 in either direction. The ionization parameters of the three wind components increased from the baseline values to $\log \xi = 4.1^{+0.2}_{-0.1}$, $3.41^{+0.05}_{-0.02}$, and $2.59^{+0.02}_{-0.05}~\mathrm{erg~cm~s^{-1}}$ when the  ``\texttt{$bb$}'' normalization is increased; and decreased to $\log \xi = 3.2^{+0.1}_{-0.1}$, $3.01^{+0.05}_{-0.03}$, and $2.13^{+0.03}_{-0.02}~\mathrm{erg~cm~s^{-1}}$ when the ``\texttt{$bb$}'' Norm is decreased. Other parameters remained nearly unchanged. These results indicate that an order-of-magnitude change in the blackbody strength produces a comparable variation in the ionization state of the gas, emphasizing the importance of including this component. As an independent check, we examined the archival HST G130M spectrum of Mrk 877, which shows a flux density of $F_{1350} = 5\times 10^{-14}~{\rm erg}~{\rm cm}^{-2}~{\rm s^{-1}}$\AA$^{-1}$ at $\lambda = 1350$~\AA. Comparing this with NGC 4151’s 2002 HST observations ($F_{1350} = 7.5\times 10^{-13}~{\rm erg}~{\rm cm}^{-2}~{\rm s^{-1}}$\AA$^{-1}$; \citealt{Xiang_2025, Kraemer_2006}), and accounting for the distance difference (NGC 4151 being 33 times closer), Mrk 877's ionizing luminosity should be roughly $5\times 10^{-14}/7.5\times 10^{-13} \times 33^2 = 73$ times higher. Given the best-fit ``\texttt{$bb$}'' model for NGC 4151 with an ionizing luminosity of $3\times 10^{43}~\mathrm{erg~s^{-1}}$, this scaling implies the ionizing luminosity of $\sim 2 \times 10^{45}~\mathrm{erg~s^{-1}}$ for Mrk 877, consistent with the current setup for the 'bb' components in SPEX. This agreement supports the adopted normalization and validates the inclusion of the blackbody component in the spectral modeling.

\subsection{Derived Properties: Launching radius, Energetics} \label{sec:derivedprop}

The key derived properties of the wind components, including their launching radii, the density profiles, mass outflow rates, and kinetic powers, are listed in Table \ref{table:parameters}.

The launching radius of the winds can be estimated using two approaches. The first method provides an upper limit, based on the assumption that the radial thickness of the absorber, $\Delta r$, does not exceed its distance from the central source. Given the relation $N_H = n\Delta r \leq f_\mathrm{v}nr$, and the definition of the ionization parameter, the maximum possible launching radius becomes:
\begin{equation} \label{eq: r_max}
    r_{\mathrm{max}} =  \frac{L_\mathrm{ion} f_\mathrm{v}}{\xi N_\mathrm{H}} = r_1 f_\mathrm{v},
\end{equation}
where $f_\mathrm{v}$ is the volume filling factor (assumed to be unity for homogeneous and volumetrically thick outflows), and $r_1 = L_{\mathrm{ion}} / \xi N_H$ is the ``absorption radius'', which can be directly computed from the best-fit values of $N_H$ and $\xi$. The absorption radius serves as an upper limit on the distance of the wind component from the ionizing source.

The second method provides a lower limit by equating the observed outflow velocity with the local escape velocity:
\begin{equation} \label{eq: r_min}
    r_{\mathrm{min}} = 2\frac{GM}{v^2_{\mathrm{out}}} = 2\frac{GM}{v^2_{\mathrm{z}}} \cos^2{\theta} = r_2 \cos^2{\theta},
\end{equation}
where $r_2 = 2GM/v^2_{\mathrm{z}}$ is the ``wind radius'', which depends directly on the best-fit value of the LOS velocity and assuming inclination angle $\theta$ of the outflow stream with respect to the LOS. Assuming a small inclination angle ($\cos{\theta} \sim 1$), the lower limit on the volume filling factor can be estimated as $f_{\mathrm{v_{min}}} =  r_2 / r_1$, subject to the condition of $r_{max} \geq r_{min}$. The wind radius, however, does not represent a strict lower limit on the actual launching radius, since the true outflow velocity could be higher than the LOS velocity depending on the geometry. A non-zero inclination angle, $\theta$, would allow the wind to be launched at smaller radii than the inferred wind radius. Nevertheless, this method offers a valuable constraint on how close to the central engine the wind can originate.

The inferred absorber radii for the three wind components span nearly three orders of magnitude from $r_1 = 5.1^{+0.9}_{-0.9} \times 10^3$, $3.1^{+0.3}_{-0.2} \times 10^4$, to $1.8^{+0.1}_{-0.2} \times 10^6$, respectively, in units of $GM/c^2$. The values of the wind radius are $r_2 = 180^{+15}_{-17}$, $1350^{+330}_{-250}$, and $690^{+140}_{-110}$ ($GM/c^2$). The overlapping ranges of their inferred radii suggest that all three outflows could potentially originate from the same general region. The limits for the volume filling factors are $0.03^{+0.01}_{-0.01}$, $0.04^{+0.01}_{-0.01}$, and $3.8^{+1.1}_{-0.8} \times 10^{-4}$. Interestingly, $pion_{\#3}$, with the lowest ionization parameter, has the lowest volume filling factor. This is different from the trend found in the highly ionized wind ($\log{\xi} > 3$) of NGC 4151 \citep{Xiang_2025}, where faster outflows tend to be more stratified and clumpy. The soft X-ray outflows do not follow a similar trend. It is possible that the $pion_{\#3}$ is part of a multi-phase medium, as the other hard X-ray outflows, possibly associated with $pion_{\#2}$, since they have the same outflow velocities. 

The mass outflow rate is computed as:
\begin{equation} \label{eq:M_dot}
    \dot M_{\mathrm{out}} = 4 \pi f_{\mathrm{cov}}\mu m_p \frac{L_{ion}^{in}}{\xi}v_{\mathrm{out}}f_\mathrm{v}
\end{equation}
where $\mu = 1.23$ is the mean atomic weight, $m_p$ is the proton mass. Our model makes it possible to calculate the ionizing luminosity seen by each component ($L_{\mathrm{ion}}^{in}$) using the values of ``\texttt{lixi}'', defined as $\texttt{lixi} = L_{ion}^{in}/{\xi}$, output from ``\texttt{$pion$}'' by its corresponding value of $\xi$. These values are listed in Table \ref{table:parameters}. Assuming unity volume filling factors, the mass outflow rates for the three components are $880^{+80}_{-68}$, $340^{+40}_{-160}$, and $2353^{+260}_{-210}$ ($M_{\odot}$/yr). These far exceed the Eddington mass accretion rate $\dot M_\mathrm{Edd} \equiv L_\mathrm{Edd}/ \eta c^2 = 10~(M_\odot / \rm yr)$, for radiative accretion efficiency of $\eta = 0.1$ and Eddington luminosity of $L_\mathrm{Edd} \equiv 4\pi GM_{BH}m_pc/\sigma_T \simeq 6\times 10^{46}~(\mathrm{erg~s^{-1}})$. Even after correcting for the minimum volume filling factors, the inferred mass outflow rates remain substantial: $32^{+3}_{-2}$, $15^{+2}_{-7}$, and $1^{+0.1}_{-0.1}$ ($M_{\odot}$/yr), with the two highly ionized wind component $pion_{\#1}$ and $pion_{\#2}$ posses the mass outflow rates comparable to the Eddington mass accretion rate. This suggests that either the UFO phase is short-lived, extremely clumpy, or additional mechanisms, likely failed winds, must be in play to maintain a coherent accretion flow.

The kinetic power of a wind, assuming it reaches a terminal velocity without significant acceleration, is given by:

\begin{equation} \label{eq:E_k}
    \dot E_k = \frac{1}{2} \dot M_{\mathrm{out}} v_{\mathrm{out}}^2
\end{equation}

For unity filling factors, three wind outflows has kinetic power of $27^{+3}_{-3}$, $1.4^{+0.4}_{-0.7}$, and $21^{+5}_{-4} \times10^{46}$ $\mathrm{erg~s^{-1}}$, which corresponds to $\dot E_k/L_{\mathrm{Edd}} = 4.6^{+0.6}_{-0.6}$, $0.25^{+0.07}_{-0.13}$, and $3.6^{+0.8}_{-0.6}$, respectively, beyond the required threshold ($\sim 5\%$) as a potential mode of AGN feedback. Even in the most conservative case with minimum volume filling factors, two higher ionized outflows still exceed $0.5~\%$ of the Eddington luminosity that is generally sufficient to initiate feedback \citep{Hopkins_2010}, and the innermost wind component $pion_{\#1}$ exceeds the $5\%~L_{Edd}$ threshold, making it a strong candidate for strong, galaxy-scale feedback \citep{Di_2005}.

To further investigate the energetics of the outflows compared to the radiation force, we calculate the outflow momentum rate $\dot p_{\mathrm{out}}$ and the radiation momentum flux $\dot p_{\mathrm{rad}}$, defined as below:
\begin{equation} \label{eq:P_out}
    \dot p_{\mathrm{out}} = \dot M_{\mathrm{out}} v_{\mathrm{out}}
\end{equation}

\begin{equation} \label{eq:P_rad}
    \dot p_{\mathrm{rad}} = \frac{L_{bol}}{c}
\end{equation}

The bolometric luminosity $L_{bol}$ seen by each wind component is estimated by $L_{bol} = 2L_{\mathrm{ion}}^{in} = 6 \times 10^{45}~\mathrm{erg~s^{-1}}$, where $L_{\mathrm{ion}}^{in}$ is the ionization luminosity seen by each component. This corresponds to an Eddington ratio of $0.1$. The derived bolometric luminosity is about three times higher than that inferred from optical measurements \citep{McLure_2004, Ho_Kim_2014}. However, our estimate is based on the UV continuum, which traces the region of peak emission from the accretion disk. Also, given the different wavelength ranges, variability, and systematic uncertainties in bolometric corrections, this discrepancy is within acceptable measurement uncertainties.

\begin{figure}
    \centering
    \includegraphics[width=0.48\textwidth]{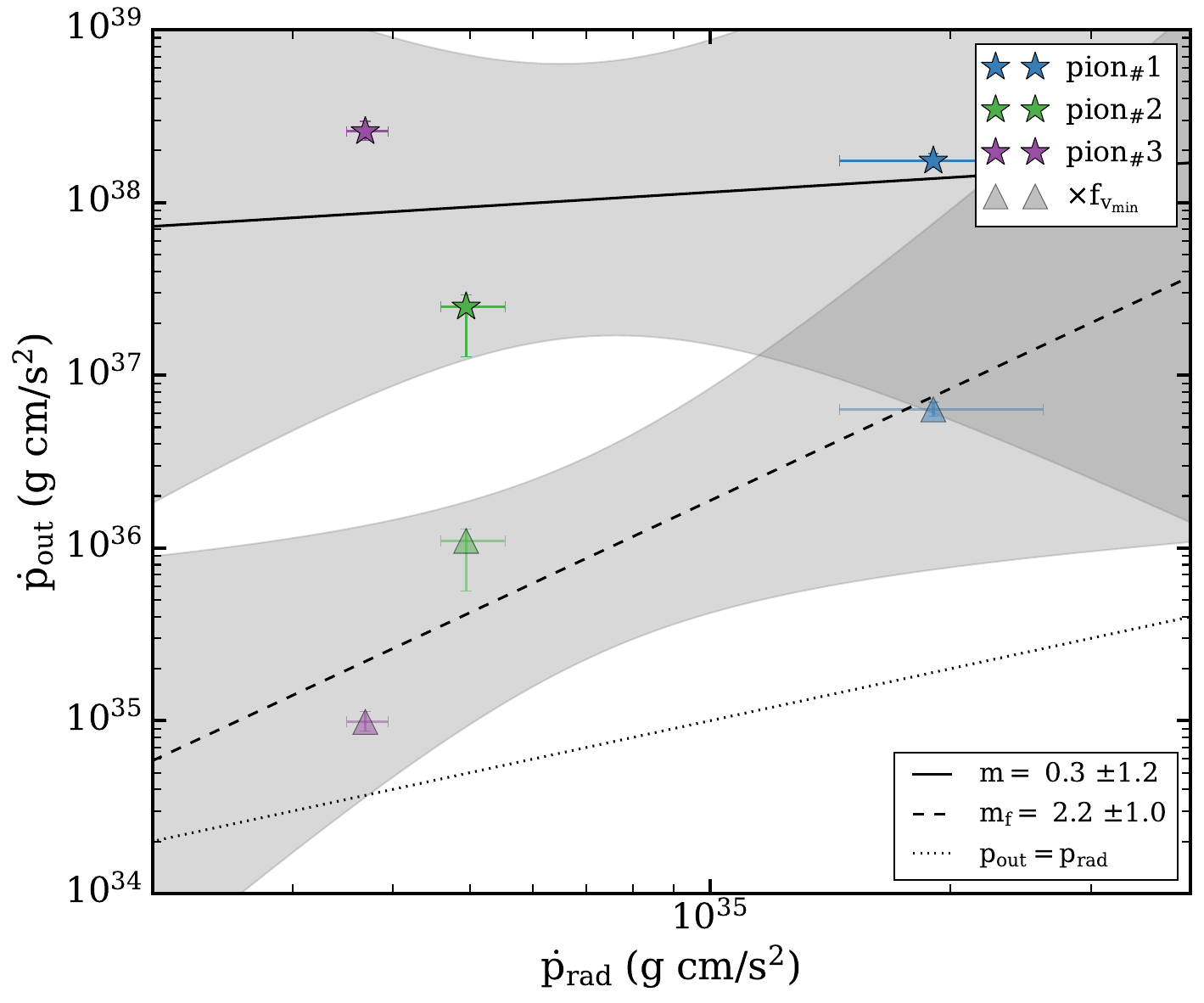}
    \caption{The outflow momentum rate ($\dot{p}_{out}$) as a function of the momentum flux of the radiation field ($\dot{p}_{rad}$). The transparent triangles represent values multiplied by the minimum volume filling factors. The solid and dashed lines are the best fit lines for the data with unity and minimum volume filling factors. The shaded regions indicating the $1\sigma$ confidence level. The best fit slope $m$ is presented in the legend. The dotted lines marked the equality between $\dot{p}_{out}$ and $\dot{p}_{rad}$}
    \label{fig:pout_prad}
\end{figure}

Figure \ref{fig:pout_prad} shows outflow momentum rate as a function of momentum flux of the radiation field, with transparent data points for values corrected by minimum volume filling factors. Using Ordinary Least Squares (OLS) linear regression via Python's \texttt{statsmodels} module fits to the two cases (unity and minimum $f_\mathrm{v}$) give slopes of $0.3\pm1.2$ and $2.2\pm1.0$, respectively, where a direct linear proportionality is not excluded. However, the large uncertainties and scatter prevent a definitive conclusion about the radiatively driven. 

In addition, the absolute values of outflow momentum rates exceed the radiation momentum flux by 1-4 orders of magnitude, even in the most conservative scenario. This discrepancy becomes more pronounced for the more ionized and higher-velocity components, suggesting that radiation pressure alone is likely insufficient and other driving mechanisms must be in play, likely the magnetically driven, especially for components closer to the black hole.

\section{Summary and Discussion} \label{sec:discussion}
We have performed photoionization modeling on a \xmm\ observation of the Seyfert 1 AGN Mrk 877. Three distinct Ultra Fast Outflow components were revealed, with line-of-sight velocities of $0.10^{+0.005}_{-0.005}~c$ , $ 0.04^{+0.005}_{-0.004}~c$ , and $0.05^{+0.005}_{-0.004}~c$, from high- to low- ionization gas respectively. Their different ionization states produce a combination of soft and hard X-ray band absorption features. The inferred launching radii span two orders of magnitude, but their radial ranges overlap, indicating that the outflows may have a common origin. 

The absence of FeK $\alpha$ emission at 6.4 keV in the data suggests that the high-column outflowing gas is obscuring the reprocessing region. The winds are likely to originate at, or interior to, the broad line region (BLR), though limited sensitivity could also explain the non-detection of FeK $\alpha$. Future high-resolution observation is required to distinguish between these scenarios. 

The large mass outflow rates inferred from the wind components should be regarded as upper limits, as they are derived under the simplifying assumption of a steady, volume-filling wind. AGN disk winds are expected to be highly clumpy and intermittent \citep{XRISMPDS456_2025, Dannen_Proga_2020, Fukumura_2015, Dannen_Proga_2020, Mehdipour_2025}.  Such clumpiness and intermittency can reproduce the observed column densities while significantly reducing the total mass carried by the flow. If the fastest wind originates from radii of order $\sim200~r_g$, the gravitational energy release per unit mass would be smaller than in the innermost disk, naturally reducing the contribution from the hottest UV-emitting regions. Theoretical models further show that strong mass loss from the disk can truncate the inner flow and suppress the far-UV emission \citep{Laor_Davis_2014}, and the integrated mass-loss rate from radiation-driven winds can even exceed the inward accretion rate.

Another caveat is that the data does not show prominent emission features (P-Cygni profiles). This is most likely attributable to the limited data quality in the current short observation. The S/N above 6 keV is modest, leading to weak constraints on any re-emission components. A similar limitation have been noted for the  FeK $\alpha$ fluorescence line, which is also undetected despite being common in many Seyfert 1 galaxies. The combination of short exposure time and the moderate resolution of the instrument makes it difficult to detect faint, spatially extended, or velocity-broadened emission features. Future deeper of higher-resolution X-ray spectroscopy will be required to better constrain the covering geometry of the outflow from the P Cygni-like features.

The fastest and most ionized component ($pion_{\#1}$), launched closest to the black hole, carries sufficient kinetic power ($\dot{K}_E > 5\%~L_{Edd}$) to drive galaxy-scale feedback even under conservative assumptions with minimum volume filling factors. Consistent with this, Mrk 877 shows a low star formation to AGN ratio in a mid-infrared diagnostics \citep{Sani_2010}, suggestive of suppressed star formation likely influenced by AGN feedback.

Absorption Measurement Distributions (AMD) suggests a radial density profile with power-law index of $\alpha = 1.54 \pm 0.04$, consistent with a magnetically driven Blandford-Payne wind. While a positive correlation between the outflow momentum rate and radiation momentum flux may support a role for radiation driving wind, the momentum carried by the outflows exceeds the available radiative momentum by at least one and up to four orders of magnitude, depending on the component and filling factor. This strongly indicates that radiation pressure alone cannot account for the acceleration of these UFOs. Moreover, thermal driving is unlikely, as the maximum launching radii for all components are at least an order of magnitude smaller than the Compton radius $R_{IC} \sim 10^7$ \cite{Begelman_1983, Woods_1996}. Therefore, a hybrid mechanism involving both radiation and magnetic is suggested. Magnetic forces likely dominate in the highly ionized, innermost component, where radiation pressure on lines becomes inefficient due to a lack of line force at $\log{\xi} > 3$ \citep{Dannen_2019}, while radiation drives the slower, lower ionized components.

\begin{figure}
\includegraphics[width=0.48\textwidth]{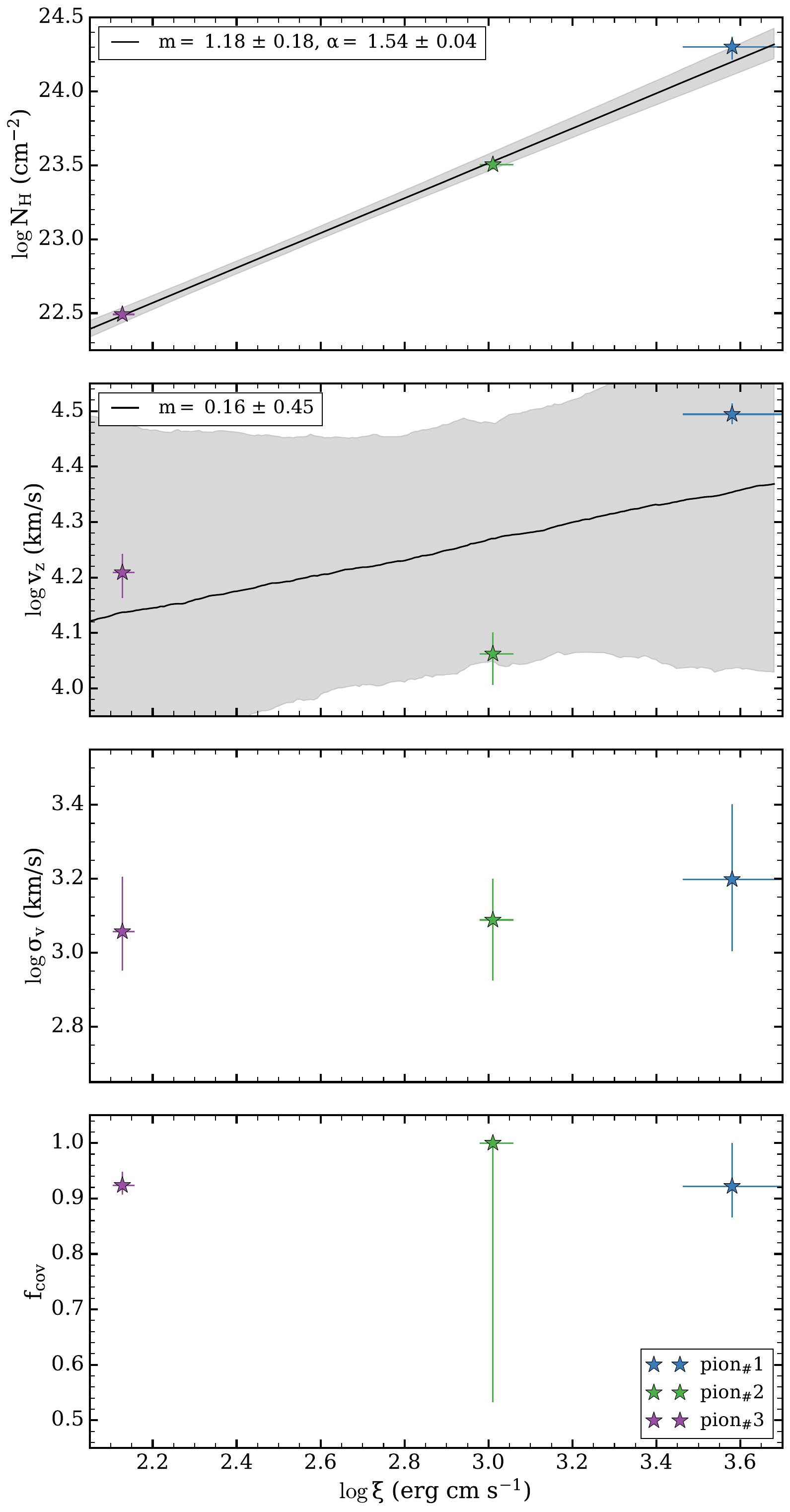}
\caption{The values of best-fit column densities ($N_H$), LOS outflow velocities ($v_z$), turbulence velocities ($\sigma_v$), and the covering factors ($f_{\mathrm{cov}}$) of the wind as a function of best-fit ionization parameter $\xi$. The best-fit line between column density and ionization parameter is shown in the top panel, with shaded regions indicating the $1\sigma$ confidence level. $m$ and $\alpha$ are the best slope and the associated density profile index with its $1\sigma$ uncertainty.}
\label{fig: propoties_vs_nH}
\end{figure}

Interestingly, the outflow velocity of the soft X-ray outflow $pion_{\#3}$ is comparable to the hard X-ray component $pion_{\#2}$, while their ionization parameters differ by 1 order of magnitude. In context of AMD with $\alpha = 1.54$, we expect the velocities correlate with ionization parameters as $v \propto \xi^{1/(4-2\alpha)}$, yielding an expected slope of $\sim 1.1$ in the $v - \xi$ plane, if the outflows arise from different radial locations that follows a relations of $v \propto r^{-1/2}$. However, in the second panel of Figure \ref{fig: propoties_vs_nH}, the OLS fits for power-law slope of the $v_z - \xi$ relation is much flatter ($m = 0.16 \pm 0.45$) than expected, mostly driven by the data point of $pion_{\#3}$. This suggests that $pion_{\#3}$ and $pion_{\#2}$ may not be radially separated, but rather from a cospatial, multiphase medium. A similar phenomenon has been observed in PG 1211+143, where a soft X-ray UFO is associated with a hard X-ray UFO component, indicative of a clumpy, two-phase wind structure \citep{Reeves_2018}. The low volume filling factor inferred for $pion_{\#3}$ supports its interpretation as a clumpier phase possibly embedded in or interacting with a higher-ionization flow.  Simulations have shown that clumping can be naturally produced in low or moderately ionized winds, particularly when the flow velocity is low enough for density perturbations to grow \citep{Dannen_Proga_2020}. As discussed above, radiation line-driving likely contributes to the soft X-ray wind component $pion_{\#3}$, helping the formation of multiphase structures, especially when shielded by failed winds. This shielding allows denser clumps to persist at lower ionization and accelerate with the same outflow stream as their high-ionization counterparts \citep{Mizumoto_2021}.

Mrk 877 is among a small number of AGNs where components of a UFO have been detected in the soft X-ray band \citep{Reeves_2018, Reeves_2003, Longinotti_2015, Gupta_2014}. The combination of similar velocities and various ionization states hints at either a failed wind or a post-shock region formed by the interaction of a high-ionization wind with circumluclear material.  Further investigation on this scenario would need a timing analysis to test the response of the properties of the absorbing gas with the continuum, which requires deeper observations and is limited by current data.

While the best-fit model is acceptable and captures the main features, several limitations remain. The modest CCD resolution may lead to multiple narrow or blended components unresolved. In addition, we did not include the reemission originating from the far side of the disk, as these features are not statistically required by the current data. Future work on deeper observations can improve on such shortcomings and may arrive at a more complete picture of the outflow. The resolution and sensitivity of XRISM \citep{Tashiro_2025, Ishisaki_2022} enable this crucial next step. Its Resolve instrument onboard delivers $\sim 4.5$ eV spectral resolution across the 1.7--12 keV band -- nearly 20 $\times$ better than CCDs like XMM-Newton/PN -- making it ideal for resolving multiple high ionization UFO components that produce Fe XXV and Fe XXVI absorption, especially those with line widths narrower than CCD resolution. Although our 18 ks XMM-Newton/PN observation of Mrk 877 revealed two such UFO components, the limited resolution ($\sim 150$ eV at 6 keV) left their line widths, driving mechanism, and structure uncertain. XRISM will not only validate these features but also uncover additional, previously unresolved wind layers, similar to the case of PDS 456 \citep{XRISMPDS456_2025}, helping to provide a more complete picture of the wind's geometry and ionization structure. These results argue for continued monitoring of Mrk 877, as it offers a unique opportunity to study the phases and variability of UFOs in an optically unobscured, moderately accreting AGN with strong indications of multiple wind layers, yet remains underexplored.

\begin{deluxetable*}{cccccc} 
\tabletypesize{\footnotesize}
\label{table:parameters}
\tablecaption{The upper panel of this table presents best-fit parameters of the model and the Detection Significance (D.S) for each outflow component. The total ionizing luminosity of the best-fit model is $L_{ion}^{all} = 2.9\times10^{45}~\mathrm{erg~s^{-1}}$. The total X-ray flux is $F_{0.3-10} = 2.9 \times 10^{-12} \mathrm{erg~cm^{-2}~s^{-1}}$ in the 0.3-10 keV band. The lower panel of this table provides the key wind parameters including launching radius ($r_1$, $r_2$), mass outflow rate $\dot M_{\mathrm{out}}$, kinetic power $\dot E_k$, the input ionizing luminosity $L_{ion}^{in}$, radiation momentum flux $\dot P_{\mathrm{rad}}$, and outflow momentum rate $\dot P_{\mathrm{out}}$.}
\tablewidth{0pt}
\tablehead{
\colhead{Parameters} &  \colhead{powerlaw} & \colhead{$pion\#1$} & \colhead{$pion\#2$} & \colhead{$pion\#3$}}
\startdata
    $\Gamma$ & $1.45^{+0.01}_{-0.01}$ & ... & ... & ... \\
    Norm ($\times 10^{52}$ $\mathrm{ph~s^{-1}~keV^{-1}}$) & $8.3^{+0.1}_{-0.1}$ & ... & ... & ... \\
    $N_H$ ($\rm cm^{-2}$) & ... & $2.0^{+0.3}_{-0.4} \times 10^{24}$ & $3.2^{+0.3}_{-0.2} \times 10^{23}$ & $3.1^{+0.2}_{-0.2} \times 10^{22}$\\
    $\log \xi$ ($\mathrm{erg~cm~s^{-1}}$) & ... & $3.58^{+0.14}_{-0.12}$ & $3.01^{+0.05}_{-0.03}$ & $2.13^{+0.03}_{-0.02}$ \\
    $\sigma_v$ ($\mathrm{km~s^{-1}}$) & ... & $1580^{+940}_{-570}$ & $1239^{+360}_{-390}$ & $1140^{+470}_{-240}$\\
    $v_z$ ($\mathrm{km~s^{-1}}$) & ... & $-31220^{+1270}_{-1450}$ & $-11540^{+1410}_{-1070}$ & $-16180^{+1620}_{-1300}$\\
    $f_{cov}$ & ... & $0.92^{+0.08}_{-0.06}$ & $1.00^{+0}_{-0.47}$ & $0.92^{+0.02}_{-0.02}$\\ 
    \hline
    $r_1$ ($GM/c^2$) Eq. \ref{eq: r_max} & ... & $5.1^{+0.9}_{-0.9} \times 10^3$ & $3.1^{+0.3}_{-0.2} \times 10^4$ & $1.8^{+0.1}_{-0.2} \times 10^6$\\
    $r_2$ ($GM/c^2$) Eq. \ref{eq: r_min}& ... & $180^{+15}_{-17}$ & $1350^{+330}_{-250}$ & $690^{+140}_{-110}$\\
    $f_{\mathrm{v_{min}}} =  r_2 / r_1$  & ... & $0.03^{+0.01}_{-0.01}$ & $0.04^{+0.01}_{-0.01}$ & $3.8^{+1.1}_{-0.8} \times 10^{-4}$ \\
    $\dot M_{\mathrm{out}}/f_\mathrm{v}$ ($M_{\odot}$/yr) Eq. \ref{eq:M_dot} & ... & $880^{+80}_{-68}$ & $340^{+40}_{-160}$ & $2530^{+260}_{-210}$ \\
    $\dot E_k/f_\mathrm{v}$ $\times 10^{46}$ ($\mathrm{erg~s^{-1}}$) Eq. \ref{eq:E_k}& ... & $27^{+3}_{-3}$ & $1.4^{+0.4}_{-0.7}$ & $21^{+5}_{-4}$ \\
    $\dot E_k/L_{\mathrm{Edd}}/f_\mathrm{v}$ & ... & $4.6^{+0.6}_{-0.6}$ & $0.25^{+0.07}_{-0.13}$ & $3.6^{+0.8}_{-0.6}$ \\
    $\dot E_k/L_{\mathrm{Edd}}/f_\mathrm{v} \times f_{\mathrm{v_{min}}}$ (\%) & ... & $16^{+2}_{-2}$ & $1.1^{+0.3}_{-0.6}$ & $0.14^{+0.03}_{-0.02}$ \\
    $L_{ion}^{in}$ ($\times 10^{45}$ $\mathrm{erg~s^{-1}}$) & ... & $2.8^{+1.1}_{-0.7}$ & $0.74^{+0.09}_{-0.05}$ & $0.55^{+0.04}_{-0.03}$\\
    $\dot p_{\mathrm{rad}}$ ($\times 10^{34}$ dyn) Eq.\ref{eq:P_rad} & ... & $19^{+0.7}_{-0.5}$ & $4.9^{+0.6}_{-0.3}$ & $3.7^{+0.3}_{-0.2}$ \\
    $\dot p_{\mathrm{out}}/f_\mathrm{v}$ ($\times 10^{38}$ dyn) Eq.\ref{eq:P_out} & ... & $1.7^{+0.2}_{-0.2}$ & $0.25^{+0.04}_{-0.1}$ & $2.6^{+0.4}_{-0.3}$ \\
    \hline 
    $\Delta AIC$ (D.S) & ... & $-37~(6.0\sigma)$& $-11~(4.3\sigma)$ & $-213~(\infty)$ \\
\enddata
\end{deluxetable*}

\begin{acknowledgments}
E.B. acknowledges support from ISF grant 2617/25. We sincerely appreciate the reviewer's thoughtful and constructive comments, which have greatly improved the clarity and physical completeness of the paper. The comments led us to refine several aspects of the analysis, including correcting the UV blackbody parameters and reassessing the corresponding ionizing luminosity. We have also expanded the discussion to address the physical plausibility of the wind scenario, particularly in reconciling the large inferred mass outflow rates and the absence of the re-emission features.
\end{acknowledgments}

\bibliography{main}{}
\bibliographystyle{aasjournal}

\end{CJK*}
\end{document}